\documentclass[a4paper, reprint, floatfix, aps, prl, superscriptaddress]{revtex4-2}

\PassOptionsToPackage{disable}{todonotes}
\usepackage{todonotes}    

\usepackage{amssymb, acronym, graphicx}
\usepackage{adjustbox}
\usepackage{lipsum}
\usepackage[normalem]{ulem}
\usepackage{graphics}
\usepackage{amsmath, amssymb, amsfonts, placeins}
\usepackage[final]{changes} 
\usepackage{color,soul}
\usepackage[hidelinks]{hyperref}
\usepackage[capitalise]{cleveref}
\usepackage[detect-all]{siunitx}
\usepackage[caption=false]{subfig}

\usepackage{textcomp}
\DeclareUnicodeCharacter{03BC}{\ensuremath{\mu}}
\DeclareUnicodeCharacter{00B5}{\textmu}

\crefname{figure}{Fig.}{Figs.}
\Crefname{figure}{Figure}{Figures}
\crefname{equation}{Eq.}{Eqs.}
\Crefname{equation}{Equation}{Equations}
\crefname{section}{Sec.}{Secs.}
\Crefname{section}{Section}{Sections}

\usepackage{placeins}
\usepackage{xcolor}
\usepackage{color,soul}

\newacro{OPO}{optical parametric oscillator}
\newacro{OPA}{optical parametric amplifier}
\newacro{SNR}{signal-to-noise ratio}
\newacro{GW}{gravitational wave}
\newacro{CLF}{coherent locking field}
\newacro{SHG}{second harmonic generator}
\newacro{PPKTP}{periodically-poled potassium titanyl phosphate}
\newacro{PDH}{Pound-Drever-Hall}
\newacro{AOM}{acousto-optic modulator}
\newacro{LO}{local oscillator}
\newacro{RIN}{relative intensity noise}

\begin{document}

\title{Loss-tolerant detection of squeezed states in the 2~\textmu m region }

\author{K. M. Kwan} 
\altaffiliation{Corresponding author}
\email{kar.kwan@anu.edu.au}
\affiliation{Australian National University, Canberra}

\author{T. G. McRae} 
\affiliation{Australian National University, Canberra}

\author{J. Qin}
\affiliation{Australian National University, Canberra}

\author{D. W. Gould} 
\affiliation{Australian National University, Canberra}

\author{S. S. Y. Chua} 
\affiliation{Australian National University, Canberra}

\author{J. Junker}
\affiliation{Australian National University, Canberra}
\affiliation{Technical University of Denmark, Kongens Lyngby}

\author{R. Iden}
\affiliation{Institute of Science Tokyo, Tokyo}

\author{V. B. Adya} 
\affiliation{KTH Royal Institute of Technology, Stockholm}

\author{M. J. Yap} 
\affiliation{Australian National University, Canberra}

\author{B. J. J. Slagmolen} 
\affiliation{Australian National University, Canberra}

\author{D. E. McClelland} 
\affiliation{Australian National University, Canberra}

\author{R. L. Ward} 
\affiliation{Australian National University, Canberra}

\begin{abstract}
Squeezed states of light enable quantum-enhanced measurements but are limited by optical loss, particularly at \SI{2}{\micro\metre} where photodiode efficiency is low. We report the first loss-tolerant, audio-band squeezed light detection at \SI{1984}{\nano\metre} by using a phase-sensitive amplifier to amplify the squeezed vacuum prior to detection. This technique increases the effective detection efficiency from \SI{74}{\percent} to \SI{95}{\percent} and increases the observed squeezing from \SI{4}{\deci\bel} to \SI{8}{\deci\bel}, the highest level of squeezing observation reported at this wavelength. Additionally, the shot–dark-noise clearance increases, extending the effective measurement bandwidth toward lower frequencies. This approach is largely wavelength-independent, extending high-fidelity quantum measurements to future gravitational-wave detectors and related quantum technologies.

\end{abstract}


\maketitle

\textit{Introduction}.--Squeezed states of light enhance sensitivity in a range of precision measurements, including \ac{GW} detection \cite{LIGO_squeezed_light_application,Virgo_squeeze,Maggie_ligosqueezing, LIGO2013_squeezing,Quantum_correlations,PhysRevLett.126.041102, Minjet_radiationpressure}, biosensing \cite{Bowen_microscopy}, and quantum information \cite{ultra_fast_computation_takanashi2020all}. Their practical utility, however, is limited by photon loss which degrades observable squeezing by introducing vacuum fluctuations and technical noise that masks the quantum correlations. While noises such as phase noise and photodiode dark noise affect all frequencies, their impact is especially pronounced in the audio band, which is relevant to \ac{GW} detection \cite{Sheila_phasenoise,sheila_thesis}. 

The transition to \SI{2}{\micro\metre} wavelength in the proposed future \ac{GW} detectors is motivated by the thermal noise reduction\cite{Rana_voyager,NEMO,Einstein_telescope}. These future detectors will employ cryogenic silicon test mass \cite{Johannes}, which has high absorption at \SI{1064}{\nano\metre}. At \SI{1064}{\nano\metre}, up to \SI{15}{\deci\bel} of tabletop squeezing has been demonstrated \cite{Vahlbruch_15dB}. Squeezing has been integrated in the kilometer-scale interferometers \cite{Maggie_ligosqueezing}, however loss and low-frequency detection remain major technical challenges.


The transition to the \SI{2}{\micro\metre} wavelength introduces new difficulties: although \SI{2}{\micro\metre} photodiodes reach quantum efficiencies as high as \SI{92}{\percent} at MHz frequencies \cite{Steinlechner_squeezedlight}, photodetectors optimized for the audio band currently achieve only \SI{74}{\percent} \cite{Georgia_squeezed_light_paper}. Reverse biasing a photodiode can improve efficiency at the expense of increased dark noise; cryogenic cooling reduces dark noise but degrades quantum efficiency \cite{Julian_photodetector_cooling}. In contrast, current \ac{GW} detectors at \SI{1064}{\nano\metre} routinely exceed \SI{98}{\percent} efficiency \cite{Maggie_ligosqueezing,Vahlbruch_15dB}. Thus, achieving high-fidelity, audio-band squeezing at \SI{2}{\micro\metre} remains a key technical hurdle for future detectors. To address the limitation of photodiode efficiency, we experimentally demonstrate quantum-enhanced detection at \SI{1984}{\nano\metre} using a phase-sensitive amplifier to mitigate the impact of low photodiode efficiency. This approach, first proposed by Caves \cite{Caves_quantumnoise,Caves_quantum_limit_in_amplifier}, enhances signal strength and preserves squeezing in the presence of detection loss. Proof-of-principle experiments have validated its loss-tolerant properties \cite{manceau_su11,Shaked_lifting_bandwidth_limit}, and integrated photonics have demonstrated broadband, room-temperature implementations \cite{fewcycle_nehra,ultra_fast_computation_takanashi2020all,Nehra_2025ultrafastallopticalmeasurementsqueezed,Furusawa_THZ_2025,Nehra_2024ultrafastsinglephotondetectionusing}. 

In the context of \ac{GW} detectors, this technique underpins internal squeezing, where in-situ amplification improves readout sensitivity \cite{PhysRevA.85.023815}. Both theoretical and experimental work show that operating in the amplification regime can surpass the standard sensitivity–bandwidth limit \cite{Vaishali_internalsqueezing,Kentaro,kentarostiffbar,Korobkoquantumexpander,korobko2023mitigating}. These developments form the basis for our demonstration of loss-tolerant quantum sensing at \SI{2}{\micro\metre}.

In this Letter, we demonstrate a loss-and-phase-noise-tolerant measurement scheme that enhances detection in the presence of limiting dark noise. By externally amplifying squeezed vacuum at \SI{1984}{\nano\metre}, we recover the squeezing degraded by low detector efficiency. Our results agree well with the model and validate the effective efficiency metric introduced to quantify detection performance \cite{Kwan_2024}. Crucially, phase noise from the \ac{OPA} has a negligible effect on observed squeezing, even at low frequencies. These findings establish a viable, loss-tolerant strategy for quantum-enhanced detection at \SI{2}{\micro\metre}—directly applicable to next-generation \ac{GW} observatories.

\begin{figure}[t!]
\centering
\includegraphics[width=\columnwidth]{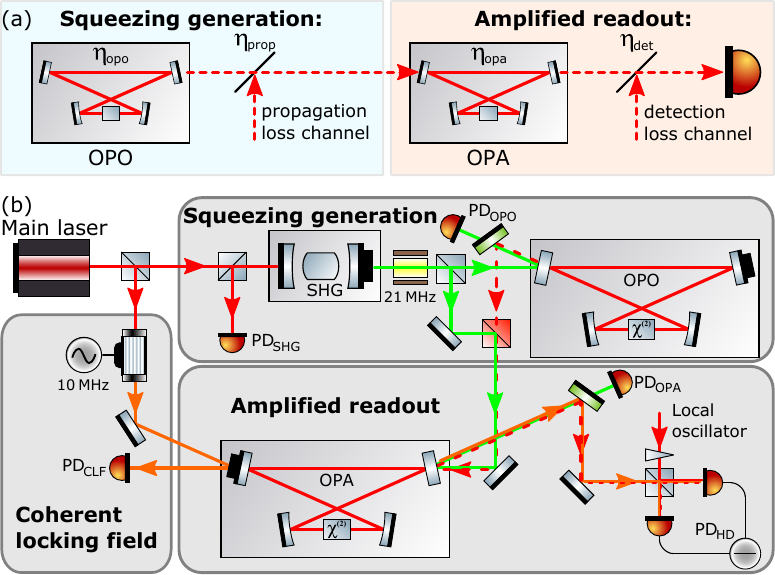}	
\caption{(a) The nomenclature for the experimental setup and the vacuum coupled into the system as loss at key points. (b) The schematic for the experiment. The squeezed vacuum is generated from a squeezing setup that consists of a \ac{SHG} and an \ac{OPO}. The amplified readout consists of an \ac{OPA} and a balanced homodyne detector. The photodiodes (PD) indicated are employed in the feedback loops. The \SI{21}{\mega\hertz} sideband is used for \ac{PDH} locking of \ac{OPO} and \ac{OPA} cavities. The \SI{10}{\mega\hertz} \ac{CLF} is used to control the pump phase of the \ac{OPA}, shown as a single AOM for simplicity. The \ac{LO}, split off from the main laser, is constrained to \SI{150}{\micro\watt} of power to avoid the photodetector saturation.}
\label{figure schematic}
\end{figure}

\textit{Experiments}.--The conceptual diagram for the squeezed vacuum generation and amplifier measurement setup is illustrated in
\cref{figure schematic}(a). A sub-threshold \ac{OPO} with an escape efficiency $\eta_{\rm opo}$ is used to generate the squeezed vacuum. Before reaching the \ac{OPA}, the squeezing encounters propagation loss, $\eta_{\rm prop}$, which in our setup is primarily due to mode mismatch between the two optical cavities. The squeezed vacuum is then amplified by an \ac{OPA} that has escape efficiency $\eta_{\rm opa}$. All losses after the \ac{OPA}, including those due to diode quantum inefficiency, are characterized by the photodetection loss $\eta_{\rm det}$. 

\Cref{figure schematic}(b) shows a simplified layout of the experimental implementation. The \ac{OPO} and \ac{OPA} are dual-resonant bow-tie cavities with an identical design, as described in \cite{Georgia_squeezed_light_paper}, with the input coupler reflectivity changed to \SI{97}{\percent} to reduce the threshold power for each cavity. A \SI{1984}{\nano\metre} thulium fiber laser from AdValue Photonics \cite{AdValue_seed} is used to provide a \ac{LO} for homodyne detection and to pump a \ac{SHG}. The \ac{SHG} is a \ac{PPKTP} crystal placed in a \SI{33}{\milli \metre} Fabry-Perot cavity that is stabilized on reflection with the Hansch-Couillaud method \cite{HC_lock}. The pump field is split into two paths to pump the \ac{OPO} and \ac{OPA}. These cavities were \ac{PDH} locked to the pump with a phase modulation of \SI{21}{\mega\hertz}. To control the pump phase of the \ac{OPA}, we use the \ac{CLF} technique where we inject a weak frequency-shifted field into the \ac{OPA} \cite{Minjet_phasecontrol,vahlbruch_clf,Chelkowski_clf}. The \ac{CLF} \SI{10}{\mega\hertz} frequency shift is generated from two \ac{AOM}'s (shown as one in \cref{figure schematic}) and is used to control the phase of the light relative to the \ac{OPA} \cite{Oelker_phase}. The \ac{LO} is locked to the \ac{CLF} using an error signal generated from the homodyne signal. This allows us to control the phase of the amplified readout scheme before injecting the squeezed vacuum into the \ac{OPA}.

\begin{table}[t!]
    \centering
    \begin{tabular}{c|c}        
        \textbf{Parameter} & \textbf{Value} \\
        \hline
        escape efficiency of OPO, $\eta_{\rm opo}$ & \SI[separate-uncertainty = true]{98.2(0.5)}{\percent} \\

        escape efficiency of OPA, $\eta_{\rm opa}$ & \SI[separate-uncertainty = true]{97.3(0.5)}{\percent} \\

        homodyne detector fringe visibility, $\eta_{\rm hd}$ & \SI[separate-uncertainty = true]{98(1)}{\percent} \\

        other propagation loss & $<$ \SI{1}{\percent} \\

        mode mismatch between \ac{OPO} and \ac{OPA} &  \SI[separate-uncertainty = true]{97(1)}{\percent}  \\

       photodiode quantum efficiency, $\eta_{\rm det}$ & \SI[separate-uncertainty = true]{74(3)}{\percent} \\

       Squeezing phase noise (without OPA), $\theta_{\rm opo}$ & \SI[separate-uncertainty = true]{46(4)}{\milli \radian} \\
       
       Squeezing phase noise (with OPA), $\theta_{\rm opo}$ & \SI[separate-uncertainty = true]{33(1)}{\milli \radian}  \\

       Amplifier phase noise, $\theta_{\rm opa}$ & \SI[separate-uncertainty = true]{218(1)}{\milli \radian} \\

    \end{tabular}    
    \caption{The experimental parameters.}
    \label{parameters table}
\end{table}

\textit{Characterizing the OPO and the OPA}.--
We measure the squeezing generated separately by the \ac{OPO} and \ac{OPA} to characterize the phase noise and the photodetector loss. For the \ac{OPO} squeezing measurement, the \ac{OPA} is bypassed. These squeezing measurements (not shown) allow a measurement of total loss and phase noise in our setup. The squeezing/antisqueezing variance $(V_{(\mp)})$ is given by
\begin{equation}
V_{(\mp)} = 1 \mp \frac{4 x \eta_{\rm tot} } {(1 \pm x)^2} ,
\label{1 opo squeezing variance}
\end{equation}
where ${\eta_{\rm tot} = \eta_{\rm opo} \eta_{\rm hd} \eta_{\rm det} }$ is the total efficiency of the setup. The normalized pump parameter $x$, often measured from the nonlinear gain $G$ is given by
\begin{equation}
x = \sqrt{\frac{P}{P_{\rm thresh}}} =  1 - \frac{1}{\sqrt{G}} ,
\label{eq: nonlinear gain}
\end{equation}
where $P$ is the pump power and $P_{thresh}$ is the threshold pump power of the cavity.

The effect of the phase noise on the squeezed state is given by 
\begin{equation} 
V_{(\mp)}(\theta_i) = V_{(\mp)}\cos^2{\theta_i} + V_{(\pm)}\sin^2{\theta_i} ,
\label{phase noise variance for single opo}
\end{equation}
where $i \in \{\rm opo,opa \}$. The losses and phase noise in the experiments are characterized and summarized in \cref{parameters table}.

Our measurements show large amplifier phase noise between the homodyne detection angle and the \ac{OPA}. This originates from the phase lock between the \ac{OPA}'s \ac{CLF} and the local oscillator. The lower \ac{CLF} power would reduce the \ac{OPA} phase noise to \SI[separate-uncertainty = true]{144(16)}{\milli \radian}, at the expense of the lock stability, due to the limited signal-to-noise ratio of \SI{2}{\micro\metre} commercial photodetector. While such excess phase noise is detrimental to measurements in the squeezed quadrature, the \ac{OPA} is operated in the anti-squeezed (amplification) quadrature, where the effect of phase noise is strongly suppressed. This effect has been modeled and verified in our previous work \cite{Kwan_2024}, confirming that amplifier phase noise contributes negligibly in the amplification regime.

\begin{figure}[t!]
\centering
\includegraphics[width=\columnwidth]{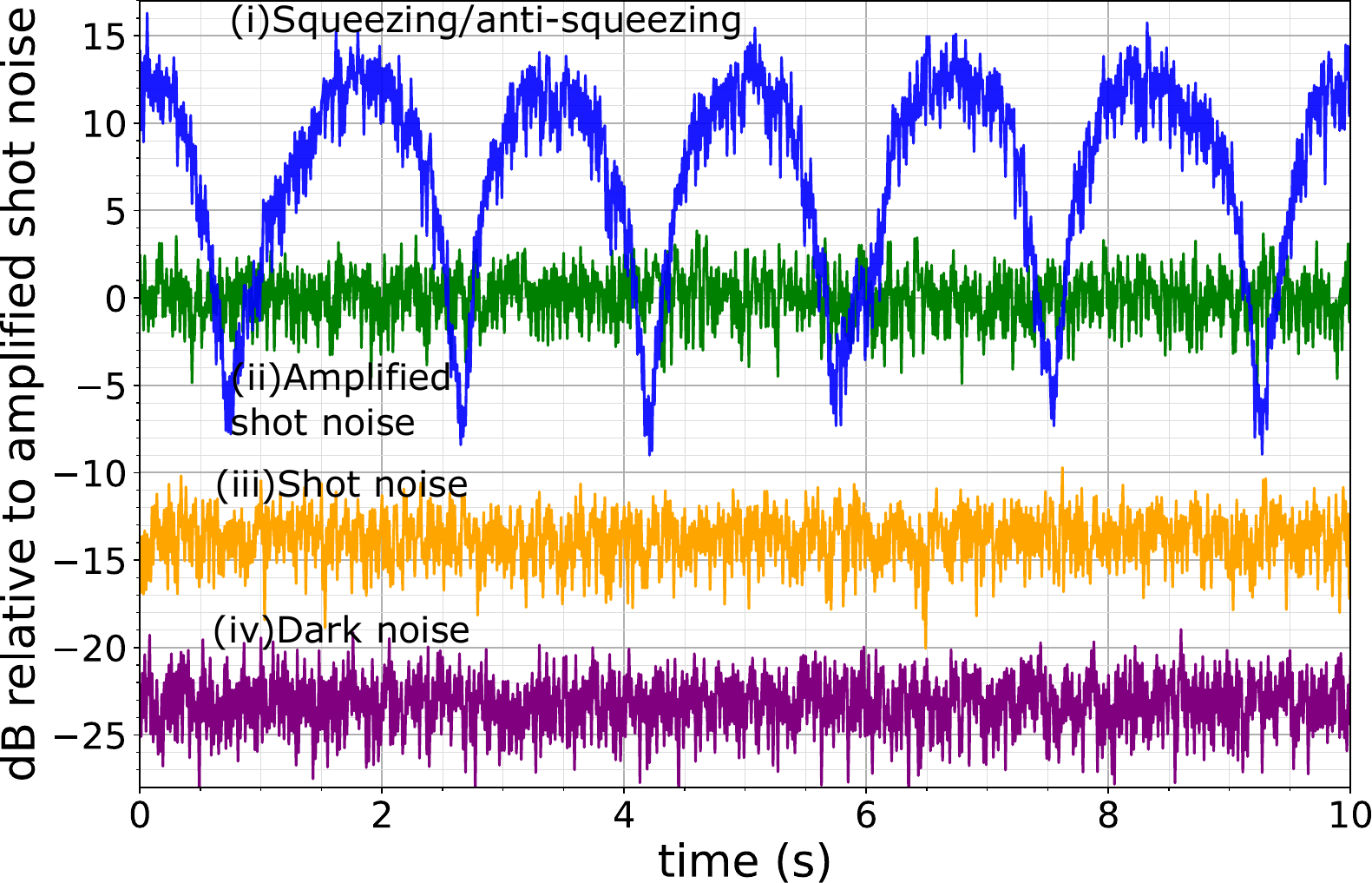}	
\caption{Measurement of the amplified squeezing level normalized to the amplified shot noise centered at \SI{52}{\kilo\hertz}. Traces correspond to: (i) amplified squeezing and antisqueezing, (ii) amplified shot noise, (iii) shot noise, and (iv) dark noise. The data is taken for $G_{\rm opo}=10$ and $G_{\rm opa}=12$, respectively. The zero span measurement was taken with a \SI{1}{\kilo\hertz} resolution bandwidth and \SI{30}{\hertz} video bandwidth.}
\label{squeezing arches}
\end{figure}

\textit{Amplified squeezing measurements}.--
We operate the \ac{OPA} locked to the antisqueezing quadrature and vary its nonlinear gain, while injecting different levels of squeezing from the \ac{OPO}. The squeezing result for $G_{\rm opo} = 10 $ and $G_{\rm opa} = 12 $ is shown in \cref{squeezing arches}. Without the \ac{OPA}, the squeezing is limited to \SI{4}{\deci\bel} with \SI{26}{\percent} of photodetection loss. With the \ac{OPA} providing \SI{14}{\deci\bel} of amplification, the observed squeezing increases to \SI[separate-uncertainty=true]{8(1)}{\deci\bel}.

\begin{figure}[b!]
\centering
\includegraphics[width=\columnwidth]{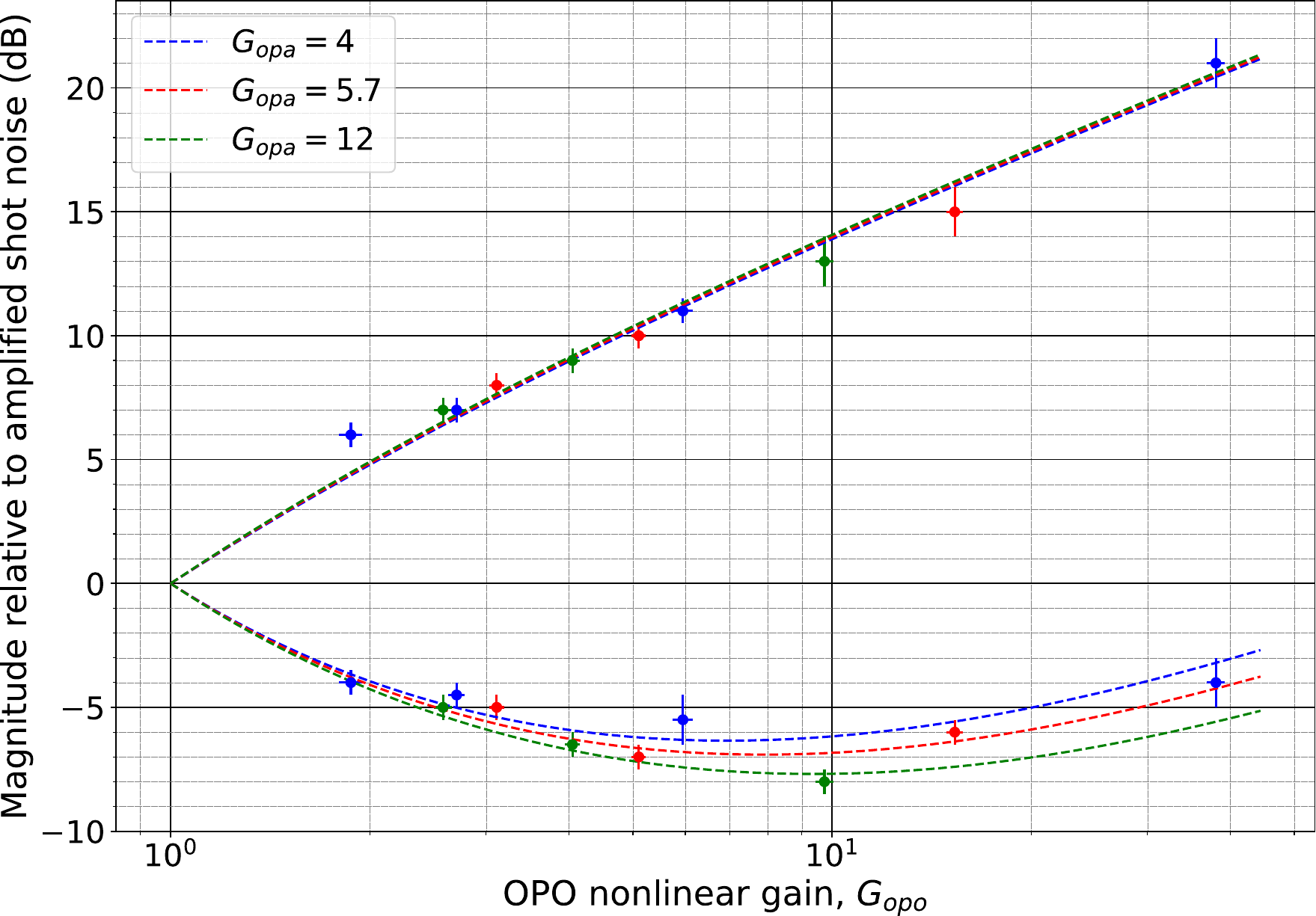}	
\caption{Amplified antisqueezing and squeezing operating versus the nonlinear gain of \ac{OPO}, with photodetector efficiency of \SI{74}{\percent} and \ac{OPA} operating at nonlinear gain of 4 (blue), 5.7 (red), and 12 (green).}
\label{OPO OPA result}
\end{figure}

The amplified squeezing and antisqueezing generated by the \ac{OPO} are shown in \cref{OPO OPA result} for the \ac{OPA} operating under various gain conditions. We fit an \ac{OPO} squeezing phase noise of \SI{33}{\milli\radian}. The \ac{OPA} with phase locking acts as a reference cavity for the \ac{OPO} and reduces the squeezing phase noise. Although the characterization of the \ac{OPA} exhibited relatively large amplification phase noise, our model predicts that \SI{218}{\milli\radian} of amplification phase noise would degrade the squeezing by less that \SI{0.5}{\deci\bel} when ${G_{\rm opo} \approx G_{\rm opa}}$. \Cref{OPO OPA result} demonstrates that amplified squeezing is largely insensitive to amplifier phase noise. In comparison, high squeezing phase noise when operating in the region of high nonlinear gain without an \ac{OPA} can significantly degrade the measured level of squeezing.

Our previous work \cite{Kwan_2024} shows that the amplified squeezing can be expressed as
\begin{equation}
V^{\rm{eff}}_{(-)} = \frac{V^{\rm amp}_{(-)}}{V^{\rm amp}_{(-)}|_{\rm G_{opo}=1}} = 1 - \frac{4 x_{\rm{opo}} \tilde{\eta}_{\rm sqz} \eta_{\rm{eff}} } {(1 + x_{\rm{opo}})^2} .
\label{equation effective variance}
\end{equation}
This equation has the same form as \cref{1 opo squeezing variance}, where $V^{\rm amp}_{(-)}$ is the amplified squeezing, $V^{\rm amp}_{(-)}|_{\rm G_{opo}=1}$ is the amplified shot noise and $\eta_{\rm eff}$ is given by
\begin{equation} 
\eta_{\rm eff} =  \frac{\eta_{\rm det}  (2\eta_{\rm opa} + x_{\rm opa} - 1 ) ^2} { (1-x_{\rm opa})^2 + 4x_{\rm opa} \eta_{\rm det} \eta_{\rm opa}  } .
\label{equation effective efficiency}
\end{equation}

Here we separate the efficiency terms into the efficiency of squeezing generation $\tilde{\eta}_{\rm sqz}=\eta_{\rm opo}\eta_{\rm prop}$ and the effective efficiency of the amplified readout $\eta_{\rm eff}$ described by \cref{equation effective efficiency}. The effective efficiency, representing the amplified readout efficiency shown in \cref{figure schematic}, is primarily tuned by varying the nonlinear gain of \ac{OPA}.

\begin{figure}[t!]
\centering
\includegraphics[width=\columnwidth]{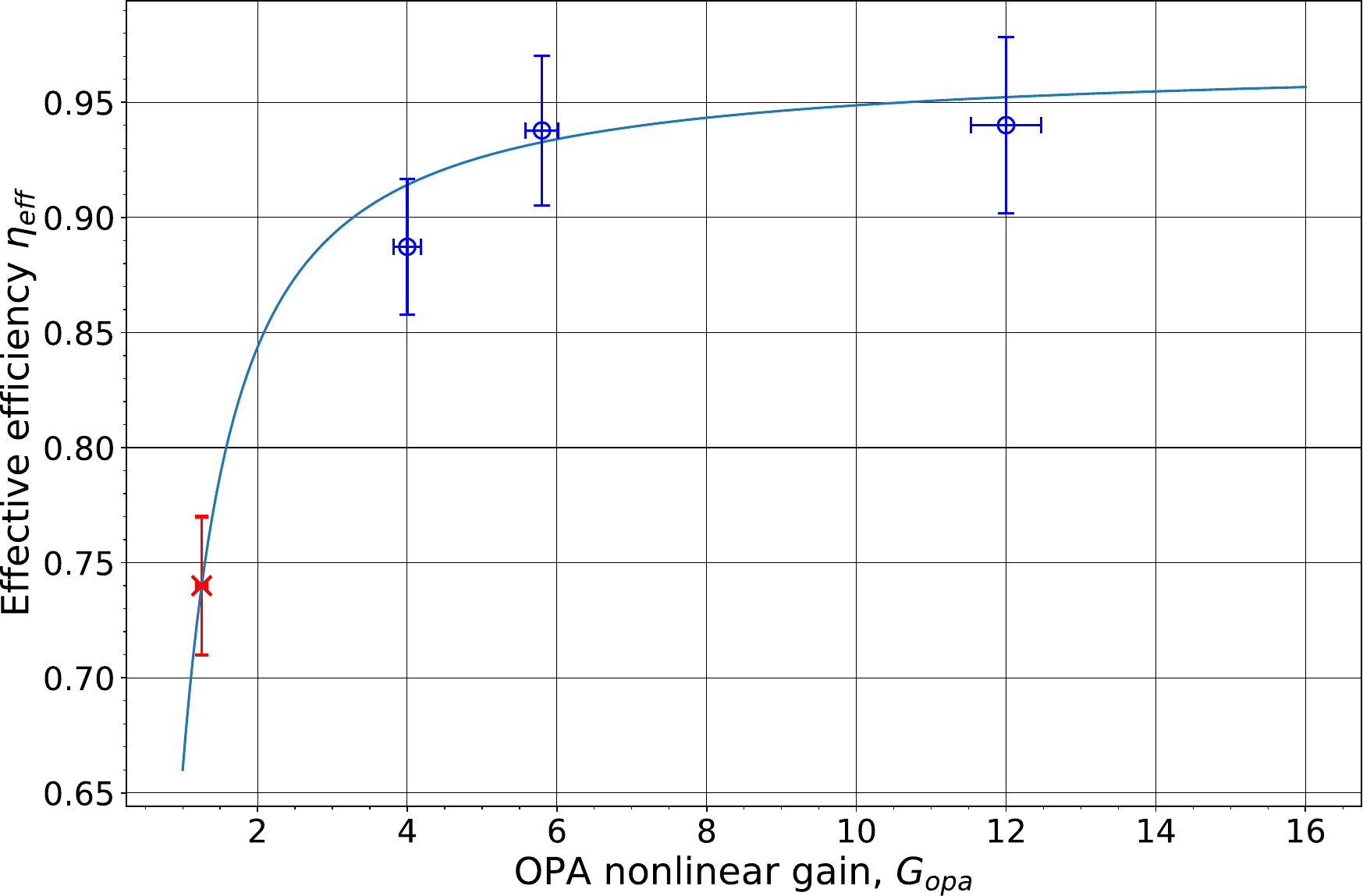}	
\caption{Enhanced effective efficiency versus the nonlinear gain of \ac{OPA}. Measured efficiencies are plotted as circles at each gain. The threshold gain where internal \ac{OPA} loss is exactly compensated is marked by a red cross. }
\label{effective efficiency plot}
\end{figure}

\Cref{effective efficiency plot} shows the enhanced efficiency of our measurement system as a function of \ac{OPA} nonlinear gain. We obtained the effective efficiency $\eta_{\rm eff}$ from our amplified squeezing and anti-squeezing results. 
The error bars for the data points in \cref{effective efficiency plot} are estimated from the fitting parameters of \cref{OPO OPA result}. A theoretical curve is included in \cref{effective efficiency plot} that is calculated from experimentally measured parameters such as photodiode quantum efficiency and \ac{OPA} escape efficiency. \Cref{effective efficiency plot} shows the increase in efficiency corresponding to the increase in \ac{OPA} nonlinear gain. The graph is plotted with the fitted $\eta_{\rm eff}$, and the $\eta_{\rm eff}$ calculated from measured $\eta_{\rm opa}$ , $\eta_{\rm det}$ and $x_{\rm opa}$ values.

We obtain the expression for the minimum pump required to compensate for \ac{OPA} internal loss $x_{\rm int}$ that can be written as
\begin{equation}
x_{\rm int} = \frac{1 - \eta_{\rm opa} }{ 1 - \eta_{\rm det}} = \frac{L_{\rm opa}}{L_{\rm det}} ,
\label{eq:internal loss gain}
\end{equation}
 where $L_{\rm opa}$ is the intracavity loss of the \ac{OPA} and $L_{\rm det}$ is the photodetection loss. This expression is obtained by setting $\eta_{\rm eff} = \eta_{\rm det}$ in \cref{equation effective efficiency}. 
Without amplification from the \ac{OPA} ($G_{\rm opa} =1$), the overall efficiency is degraded due to the additional intracavity loss introduced by the \ac{OPA}. In our measurement setup, the \ac{OPA} requires a nonlinear gain of 1.25 to compensate for the \SI{2.7}{\percent} intracavity loss introduced by the amplifier highlighted with the red cross on \cref{effective efficiency plot}.  The upper limit for the amplified readout scheme is set by $x_{\rm int}$, where a lossless \ac{OPA} $(x_{\rm int} = 0)$ is required to achieve $\eta_{\rm eff} = 1$. With a modestly high gain ($G_{\rm opa} > 10$), we were able to achieve an efficiency of \SI{95}{\percent} with our \ac{OPA}.

\begin{figure}
\centering
\includegraphics[width=\columnwidth]{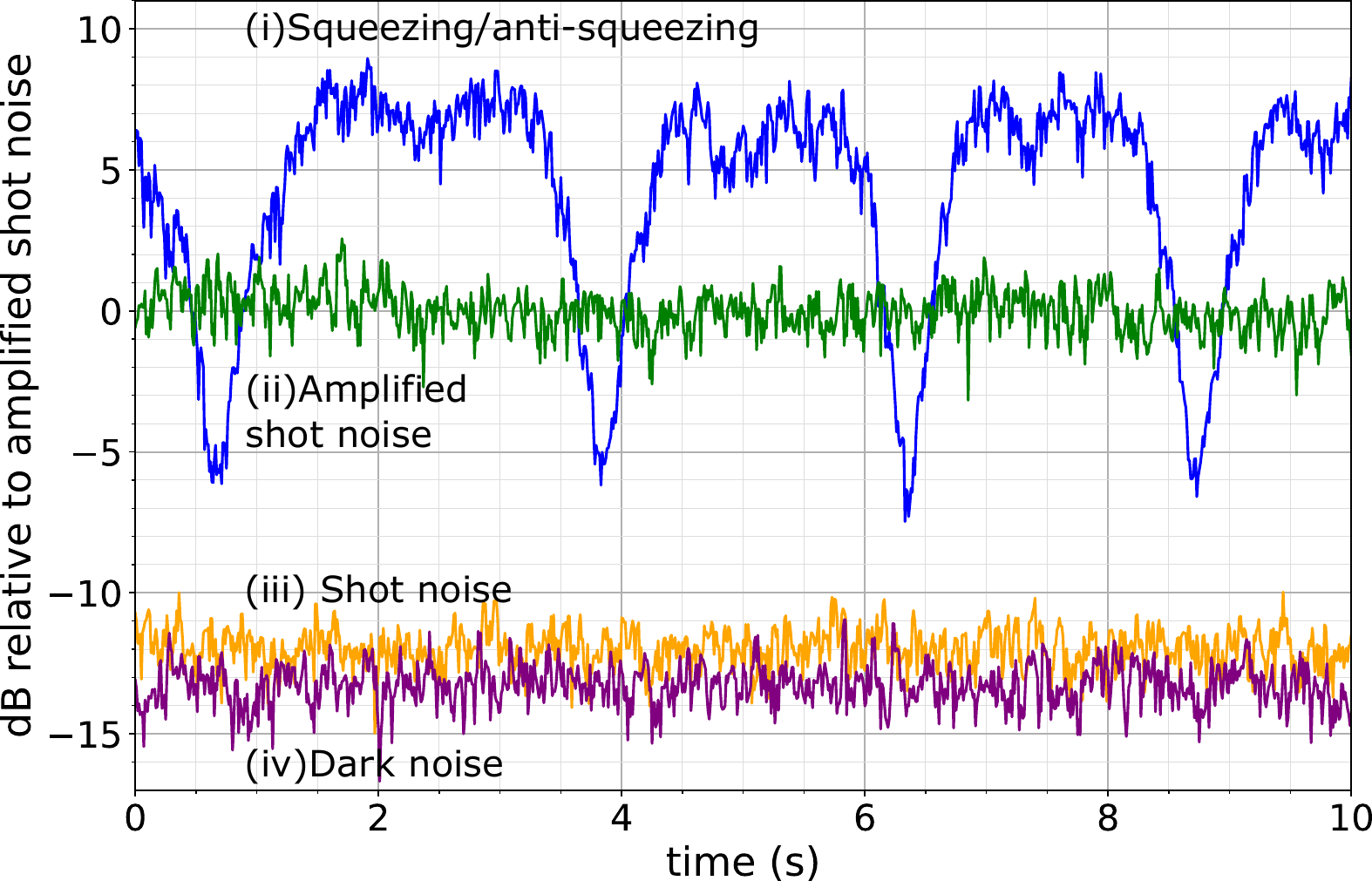}	
\caption{Measurement of the amplified squeezing at $G_{\rm opo}=3$ at \SI{3}{\kilo\hertz}. The traces correspond to: (i) amplified squeezing and antisqueezing, (ii) amplified shot noise, (iii) shot noise, and (iv) dark noise. The additional noise clearance is obtained from $G_{\rm opa}=19$. The zero span measurement was taken with a \SI{1}{\kilo\hertz} resolution bandwidth and \SI{10}{\hertz} video bandwidth.}
\label{diagram low freq}
\end{figure}

The reduction of technical noise such as electronic noise or photodiode dark noise is important at low frequencies for applications such as \ac{GW} detection where high levels of squeezing are essential. As shown in \cref{diagram low freq}, at \SI{3}{\kilo \hertz} the noise clearance is reduced to \SI{1}{\deci\bel}, where the limiting noise source originates from the electronic noise floor of the spectrum analyzer rather than from the photodiode. Conventional detection schemes typically require more than $\SI{10}{\deci\bel}$ of clearance to observe squeezing. By contrast, the \ac{OPA} adds \SI{12}{\deci\bel} of noise clearance, enabling the observation \SI{6.5}{\deci\bel} of squeezing in the noise-limited region. While the noise limitation here is electronic rather than photodiode, the same principle still holds, where phase-sensitive amplification extends the bandwidth of measurement systems.


The bandwidth extension with its upper limit set by the \ac{OPA} cavity linewidth, is especially valuable when the \ac{LO} power is limited, where further increase in \ac{LO} power can couples additional technical noise such as \ac{RIN} or Johnson noise from heating of the photodiode, limiting the observed squeezing. In this regime, some sources of dark noise, such as flicker noise that scales with $1/f$, are not well understood. By amplifying the shot noise, the \ac{OPA} can potentially mitigate the limitation of the photodiode's semiconductor properties.

\textit{Conclusions and discussions}.--We observed \SI{8}{\deci\bel} of squeezing in \SI{2}{\micro\metre} region with \SI{74}{\percent} photodetector quantum efficiency. This experiment demonstrates an alternative wavelength-independent technique to achieve high detection efficiency. At \SI{2}{\micro\metre}, we demonstrate \SI{95}{\percent} detection efficiency, among the highest reported, while extending the bandwidth into the dark-noise-dominated low frequency region. Although lower intracavity losses have been achieved, developing a photodetector with an equivalent loss figure presents a much greater challenge at this wavelength. With an \ac{OPA} that has an escape efficiency of \SI{99}{\percent} \cite{Vahlbruch_15dB}, we could improve the efficiency of the measurement scheme to \SI{98}{\percent}, a level comparable to the photodiodes currently used in the gravitational wave detectors. We have shown that phase noise in the \ac{OPA} has a minimal effect on the observation of squeezing. In addition, this technique allows amplification above dark noise in the regime limited by the semiconductor properties.
\\


\begin{acknowledgments}
\textit{Acknowledgments}.--This research was supported by the Australian Research Council under the ARC Centre of Excellence for Gravitational Wave Discovery, Grant No. CE230100016. The authors declare no competing interests. This work has been assigned LIGO document number P2500501.

R. Iden would like to acknowledge the support and funding from JST ASPIRE, Grant Number JPMJAP2320.

V. B. Adya would like to acknowledge the support and funding from the Swedish Research Council (VR starting grant 2023-0519, the Wallenberg Center for Quantum Technology (WACQT) and the Göran Gustafsson Foundation in Sweden.
\end{acknowledgments}

\bibliographystyle{apsrev4-2}
\bibliography{citation}

\end{document}